\documentclass[pss]{wiley2sp} 
\usepackage{amsmath}

\usepackage{graphicx}
\usepackage{dcolumn}
\usepackage{bm}
\newcommand{\Ru}{Ce(Ru$_{1-x}$Fe$_{x}$)PO\,}
\newcommand{\As}{CeFe(As$_{1-y}$P$_{y}$)O\,}

\begin{document}

\title{Avoided ferromagnetic quantum critical point: Antiferromagnetic ground state in substituted CeFePO}

\titlerunning{Avoided FM QCP in substituted CeFePO}

\author{%
A. Jesche\textsuperscript{\textsf{\bfseries 1,2}},
T. Ball\'e\textsuperscript{\textsf{\bfseries 3}},
K. Kliemt\textsuperscript{\textsf{\bfseries 3}},
C. Geibel\textsuperscript{\textsf{\bfseries 1}},
M. Brando\textsuperscript{\textsf{\bfseries 1}},
C. Krellner\textsuperscript{\textsf{\bfseries 1,3} \Ast}}

\authorrunning{A. Jesche et al.}

\mail{e-mail
  \textsf{krellner@physik.uni-frankfurt.de}}

\institute{%
  \textsuperscript{1}\,Max-Planck-Institute for Chemical Physics of Solids, D-01187 Dresden, Germany\\
  \textsuperscript{2}\,Center for Electronic Correlations and Magnetism, Augsburg University, D-86159 Augsburg, Germany\\
  \textsuperscript{3}\,Institute of Physics, Goethe University Frankfurt, D-60438 Frankfurt am Main, Germany}

\received{XXXX, revised XXXX, accepted XXXX} 
\published{XXXX} 


\abstract{%
%
%
%
\abstcol{%
  We have investigated single crystals of two substitution series Ce(Ru$_{1-x}$Fe$_{x}$)PO and CeFe(As$_{1-y}$P$_{y}$)O in the vicinity to the quantum critical material CeFePO by means of magnetic-susceptibility and specific-heat measurements. We observe an antiferromagnetic ground state in the vicinity of the quantum critical point, with pronounced metamagnetic transitions for $H\parallel c$, which is the magnetically hard direction. Our results verify that a ferromagnetic quantum critical point is avoided in substituted CeFePO, because we clearly demonstrate  that the ferromagnetic ground state changes into an antiferromagnetic one, when approaching the quantum critical point. 
}}

\titlefigure[width=0.8\columnwidth]{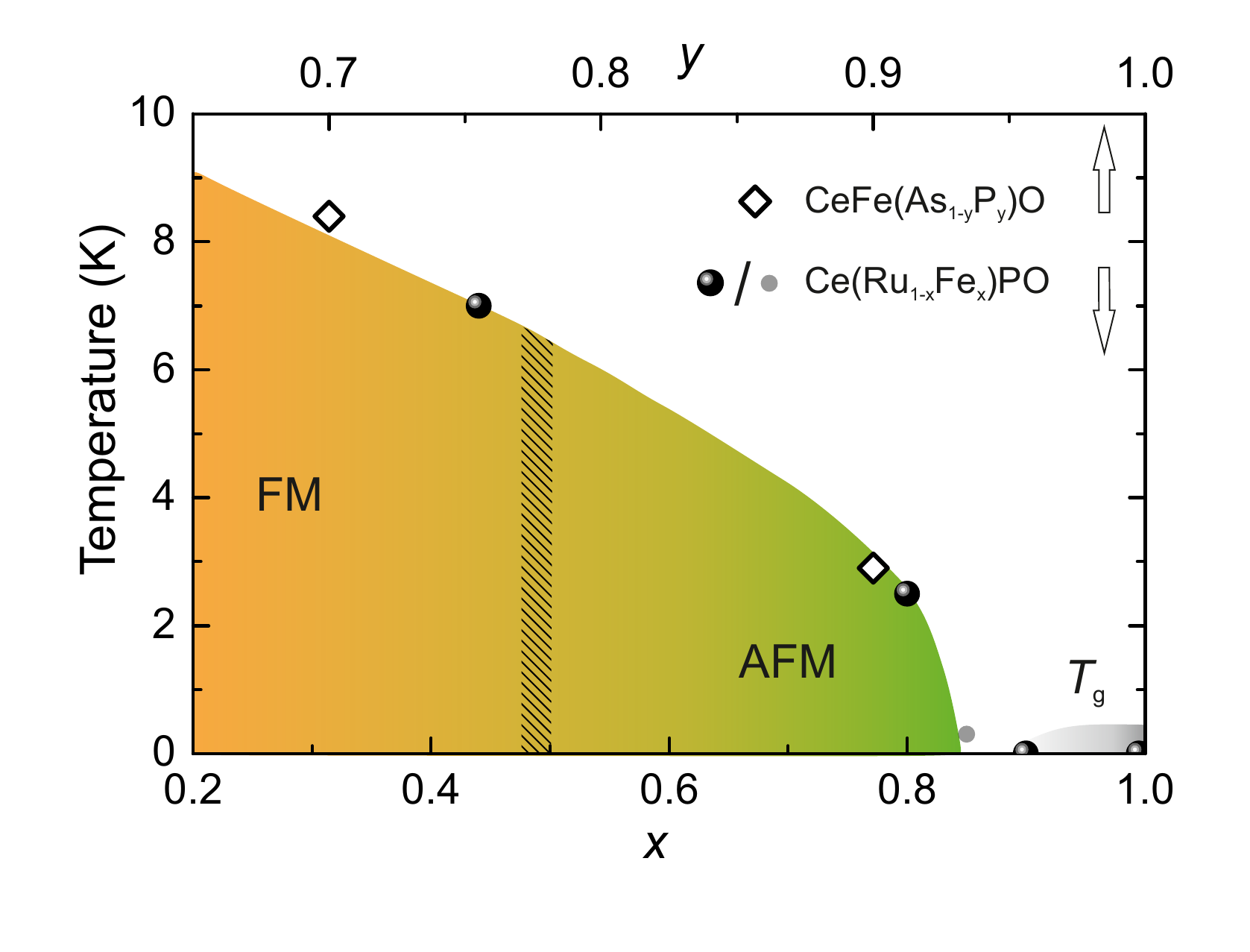}
\titlefigurecaption{%
  Temperature-substitution phase diagram of the series Ce(Ru$_{1-x}$Fe$_{x}$)PO and CeFe(As$_{1-y}$P$_{y}$)O.
}

\maketitle   

\section{\label{sec:Introduction} Introduction}

In recent years, ferromagnetic (FM) heavy-fermion metals were intensively investigated and the main question is whether a FM quantum critical point (QCP) generally exists and, if not, which are the possible ground states of matter that replace it.  Substantial experimental and theoretical efforts were made to investigate this problem, revealing a wide range of possibilities \cite{Stewart:2001,Lohneysen:2007,Brando:2016}. On theoretical grounds, it was shown that in 2D and 3D the quantum phase transition from a metallic paramagnet to an itinerant ferromagnet in the absence of quenched disorder is inherently unstable, either towards a first order phase transition or towards an inhomogeneous magnetic phase (modulated or textured structures) \cite{Belitz:2005,Conduit:2009,Kirkpatrick:2012}. The mechanism behind this phenomenon is analogous to what is known as a fluctuation-induced first-order transition in superconductors and liquid crystals \cite{Brando:2016}. 

In $4f$ systems, the theoretical description of the interplay of ferromagnetism and the local Kondo-interaction is much less explored and different scenarios are proposed \cite{Yamamoto:2010,Peters:2012}. From the experimental point of view, three main scenarios have been observed in the vicinity of a FM instability. In the first case, like in CeRu$_2$Ge$_2$ \cite{Sullow:1999}, the FM transition changes into antiferromagnetic (AFM) order, when the ordering temperature is suppressed to $T\rightarrow 0$ by e.g., hydrostatic pressure. Similar behavior was observed in CeAgSb$_2$  where a crossover to AFM order was observed at 3.5 GPa \cite{Sidorov:2003}. In the second case, like in the alloy CePd$_{1-x}$Rh$_x$ \cite{Westerkamp:2009}, it seems that local disorder-driven mechanisms such as Kondo disorder or the quantum Griffiths phase scenario are responsible for the non-Fermi liquid (NFL) properties around the QCP, but not the quantum critical fluctuations. Characteristic of this scenario are phase diagrams, where the FM transition temperature, $T_C$, smears out as function of the control parameter. A third case was recently reported, namely the clear evidence of a FM QCP in a heavy-fermion metal in arsenic-doped YbNi$_4$P$_2$, which is a quasi-1D system \cite{Krellner:2011}, \cite{Steppke:2013}. The existence of a FM QCP was also suggested in recent studies by Kitagawa \textit{et al.} \cite{Kitagawa:2012,Kitagawa:2013} in the compound \Ru with $x=0.86$, determined by NMR measurements on polycrystalline samples in finite magnetic field. 

In this article, we will present magnetization and specific-heat data measured on single crystals of this alloy series which clearly reveal that a transition to an AFM ground state is observed before the QCP is reached. 

\section{\label{sec:CeFePO} {{CeFePO} Review}}
The heavy fermion (HF) metal CeFePO has been intensively investigated due to its proximity to a magnetic quantum critical point (QCP). Characteristic of this material are strong ferromagnetic (FM) fluctuations together with  a pronounced two-dimensional anisotropy of the crystal structure. CeFePO is isoelectronic to the parent compound of iron-pnictide superconductors, CeFeAsO, and to the FM Kondo lattice CeRuPO. Zimmer \textit{et al.} synthesized polycrystalline samples of CeFePO already in 1995 \cite{Zimmer:1995}, however, no physical properties were reported at that time. From the volume plot of the lanthanide series a mixed or intermediate valence state was proposed.

Initial physical measurements on polycrystalline samples reveal a Ce$^{3+}$ state and a paramagnetic heavy-fermion ground state with a large Sommerfeld coefficient of $\gamma = 0.7$\,Jmol$^{-1}$K$^{-2}$ \cite{Bruning:2008}. Already at that time the vicinity to a FM QCP was proposed for CeFePO, concluded from a strongly field dependent and enhanced magnetic susceptibility at low temperatures. Shortly after, the occurrence of magnetic Ce$^{3+}$ was confirmed by Kamihara \textit{et al.} \cite{Kamihara:2008a} which propose that superconductivity is absent in CeFePO in contrast to the isoelectronic LaFePO, because of the magnetic Ce-moments. Subsequently, the band-structure of CeFePO and the role of the $3d$-$4f$ hybridization was theoretically determined using the local density approximation combined with dynamical mean-field theory \cite{Pourovskii:2008}. 

First single crystals were grown using a high-temper\-atu\-re Sn-flux technique \cite{Jesche:2011}, similar to what was used for the crystal growth of CeRuPO \cite{Krellner:2008b}. These single crystals were investigated with angle-resolved photoemission spectroscopy, which corroborate the sizable $3d$-$4f$ hybridization in CeFePO. Polycrystalline samples of the \As substitution series were synthesized by Luo \textit{et al.} \cite{Luo:2010}, which reveal a complex magnetic phase diagram. At the As-rich side the phase diagram is dominated by the intricate interplay between $3d$ and $4f$ moments, which leads to a variety of new phenomena for $y=0.3$ \cite{Jesche:2012}. At the P-rich side, the $d$-magnetism is vanished and the ground state is dominated by the interaction of the 4$f$-moments. There, a FM ground state of the $4f$-electrons was proposed for $0.4 \leq y \leq 0.9$, although no remanent magnetization could be resolved for the $y=0.9$ sample \cite{Luo:2010}. The same group also reported a study of the same series containing 5\% fluorine substituted on the oxygen site \cite{Luo:2011}. For CeFePO this F-doping leads to a stabilization of the magnetic phase, although the nature of the ground state could not be definitively determined, because only measurements on polycrystals down to 2\,K were reported.
The magnetic nature of Fe in CeFePO was probed by $^{57}$Fe M\"ossbauer spectroscopy down to 10\,K. No magnetic splitting was observed indicating a paramagnetic phase of the
Fe sublattice \cite{Nakamura:2012}.

Resistivity measurements under pressure on CeFePO single crystals were conducted by Zocco \textit{et al.} \cite{Zocco:2011}. They  observed a stabilization of the Kondo screening with pressure, which is reflected in an increase of the coherence temperature. This result is in agreement with the general behavior of Ce-based Kondo-lattice systems under pressure \cite{Stewart:2001}.  
A thorough $^{31}$P-NMR study on oriented powder revealed a metamagnetic transition at 4\,T, when the field was applied in the basal plane of aligned powder \cite{Kitagawa:2011}. Furthermore, Kitagawa \textit{et al.} show that around the metamagnetic transition the nuclear spin-lattice relaxation rate, $1/T_1T$, increases with decreasing temperature down to 100\,mK, a hallmark of non-Fermi liquid behavior. They propose a Kondo-breakdown scenario at this metamagnetic transition accompanied with a drastic change of the Fermi surface. Later on, a comprehensive study of the ac susceptibility, specific-heat and $\mu$SR measurements on CeFePO single crystals reveal, in contrast to the results on polycrystals, a strongly correlated short range ordered state below the freezing temperature $T_g=0.7$\,K \cite{Lausberg:2012}. This unusual short-range ordered state was ascribed to the avoidance of a FM QCP in this system. However, the strong sample dependency, the paramagnetic Fermi-liquid ground state for the polycrystals and the short range strongly correlated state for the single crystals, is reminiscent to what was observed for the prototypical CeCu$_2$Si$_2$ \cite{Modler:1995} and proves the importance of quantum fluctuations for the ground state of this system. The recent statement of the existence of a FM QCP in the series \Ru for $x=0.86$ \cite{Kitagawa:2012,Kitagawa:2013} further stimulated the interest in this layered compound and we address this issue by investigating single crystals grown in Sn-flux. 

The single crystal growth of 1111-type iron pnictide compounds is a real challenge. After the discovery of high-temperature superconductivity in F-substituted LaFeAsO in 2008 \cite{Kamihara:2008} the initial flurry of activities mainly were performed on the LnFePnO (Ln = rare earth, Pn = P, As) systems (1111), however, shortly after the focus has rapidly been shifted towards the AFe$_2$As$_2$ (A = Ba, Sr, Ca) abbreviated as 122 and FeSe/Te superconductors (11 materials), even though the latter two classes of materials have lower $T_c$ \cite{Stewart:2011}. Meanwhile, the 122 and 11 compounds are the model systems among iron-pnictide superconductors, because large and homogeneously doped single crystals can easily be achieved. However, the magnetic and electronic anisotropies are much weaker, and $T_c$ is lower in the 122 compared to the 1111 systems. In contrast, the growth of sizable high-quality single crystals of the 1111 compounds is extremely challenging, despite extensive worldwide efforts, slowing down the scientific progress in this type of compounds. Conventional solid-state reactions have been predominantly used to synthesize polycrystalline doped and undoped LnFeAsO. The main problems with the single crystal growth of the 1111 system are the following \cite{Yan:2011}: (i) The multicomponent phase diagrams are unknown. (ii) The compounds form strongly peritecticly. (iii) The presence of stable secondary phases (e.g. stable rare-earth oxides). (iv) The low solubility of oxygen in metallic and salt fluxes. In the literature, mainly the flux method was reported for the successful crystal growth of the 1111-type of materials, using NaCl/KCl, NaI, NaAs as the flux \cite{Karpinski:2009,Jesche:2012a,Yan:2009}.

\section{\label{sec:exp} {Experimental details}}
Here, we employed the high-temperature Sn-flux method, which was found to be a suitable flux for CeRuPO \cite{Krellner:2008b} and the series \As \cite{Jesche:2011}. In a first step, Pn and Sn were heated up to 600$^{\circ}$C for 5\,h in an alumina crucible which was sealed
inside an evacuated silica ampoule. In a second step, Ce, Fe, RuO$_2$, SnO$_2$, and Sn were added and the alumina crucible was sealed inside a Ta container under argon atmosphere. 
The mixture was then heated up to 1500$^{\circ}$C, slowly cooled down to 900$^{\circ}$C within one week followed by fast cooling down to room temperature. To remove the excess Sn, the samples were centrifugated at 500$^{\circ}$C and then put into diluted hydrochloric acid for 10 min. This resulted in platelike single crystals with a side length of typical 0.5\,mm but in some cases going up to more than one millimeter. 
We have grown several substitution levels of the two series \Ru and \As. The iron, $x$, and phosphorous, $y$, concentrations are given in nominal values. Energy dispersive X-ray analysis and X-ray powder diffraction set an upper limit for the relative error of the actual content $\frac{\Delta x}{x}\sim0.1$. X-ray powder diffraction patterns of ground single crystals were recorded on a Stoe diffractometer in transmission mode or a Bruker D8 diffractometer using Cu K$\alpha$-radiation. The refined lattice parameters were found to be in good agreement with the values reported in Ref.~\cite{Kitagawa:2012} for \Ru and in Ref.~\cite{Luo:2010} for \As.  Magnetic measurements were performed in a commercial SQUID VSM and the VSM option of the PPMS. Specific heat was measured with the He3-option of the PPMS. 

\begin{figure}
\includegraphics[width=0.9\columnwidth]{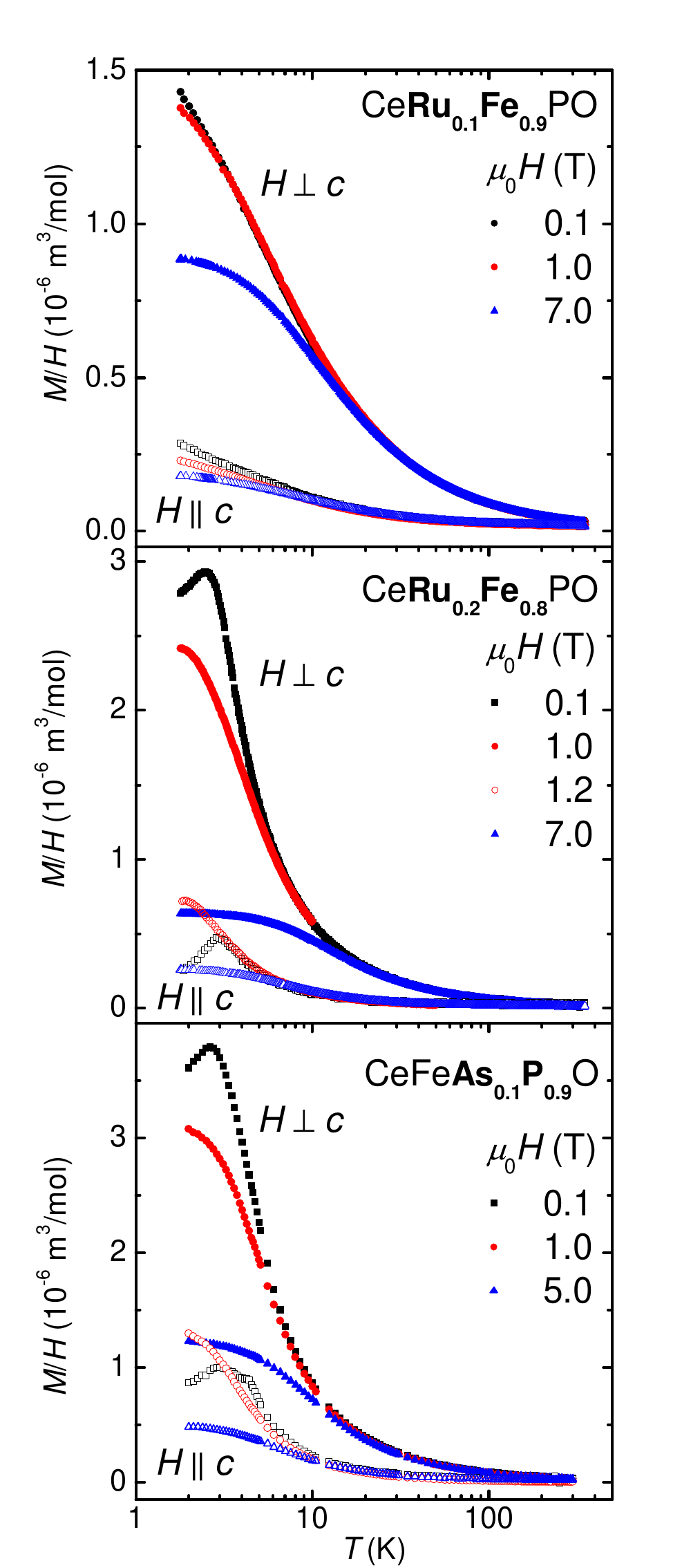}%
\caption{\label{fig1} (Color online) Temperature dependence of the magnetization divided by magnetic field (dc-susceptibility) for single crystals with $x=0.9$, $x=0.8$, and $y=0.9$. Closed (open) symbols present data measured with the magnetic field perpendicular (parallel) to the crystallographic $c$-axis.}%
\end{figure}

\section{\label{sec:phys} {Physical properties}}
\begin{figure}
\includegraphics[width=\columnwidth]{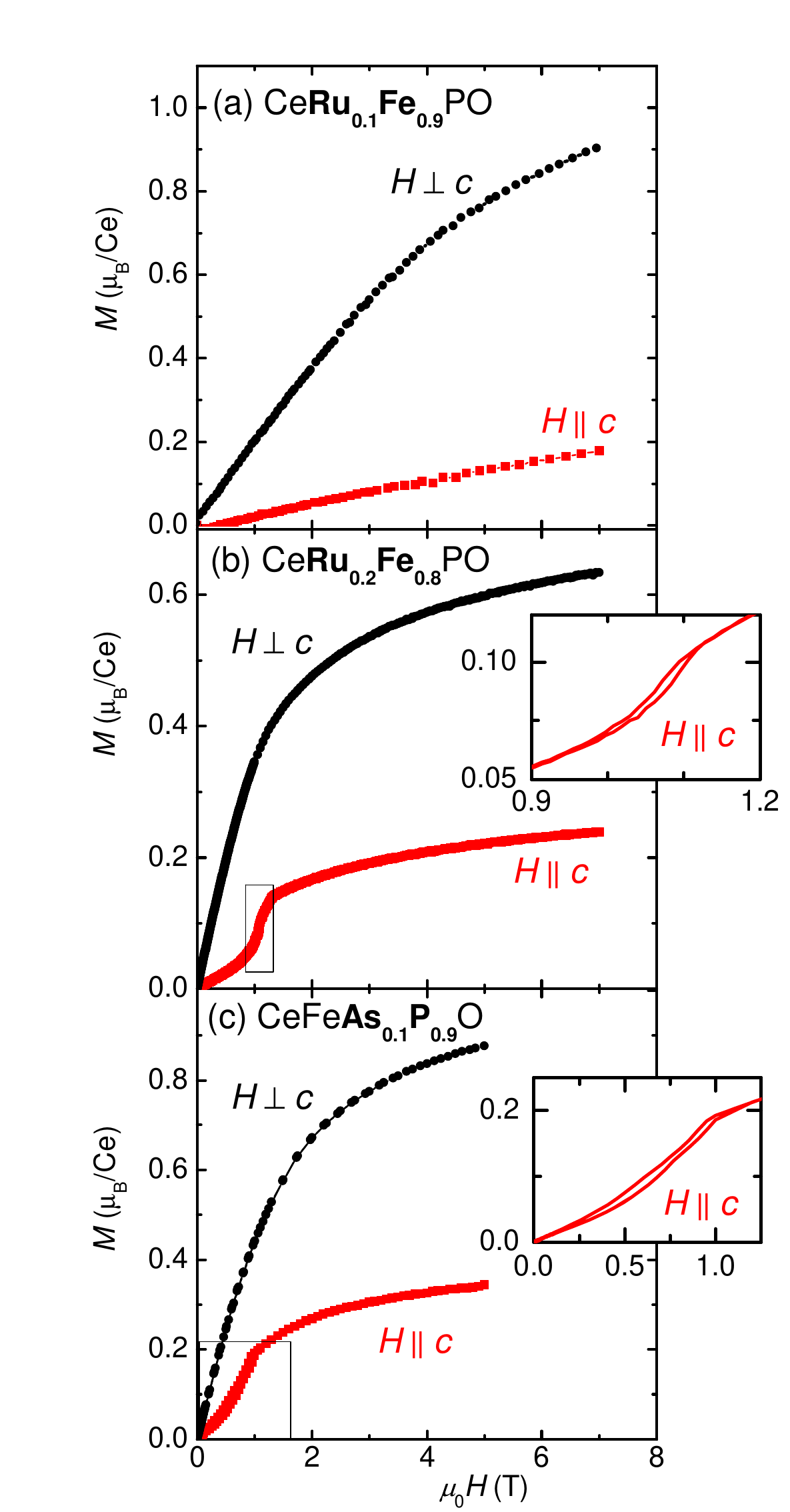}%
\caption{\label{fig2} (Color online) Magnetic-field dependence of the magnetization at $T=2$\,K for single crystals with $x=0.9$, $x=0.8$, and $y=0.9$. Black (red) points present data measured with the magnetic field perpendicular (parallel) to the crystallographic $c$-axis. In the insets for $x=0.8$ and $y=0.9$, metamagnetic behavior is visible for $H\parallel c$.}%
\end{figure}

In Fig.~\ref{fig1}, we present the temperature dependence of the  magnetization divided by magnetic field, $M/H(T)$,  for three different single crystals. For $x=0.9$, no magnetic phase transition was observed down to 1.8\,K. The magnetic anisotropy caused by the crystalline-electric field (CEF) is similar to what was observed in CeRuPO and CeFePO, i.e., the susceptibility is larger for magnetic field perpendicular to the $c$-direction (easy-plane system). For $x=0.8$ and $y=0.9$ the dc-susceptibility curves for $H\perp c$ are strongly comparable to each other. A magnetic phase transition is apparent above 2\,K, with distinct maxima at small fields. Below 10\,K a pronounced magnetic-field dependence of the $M/H(T)$ curves is observed for both field directions, with decreasing absolute values for increasing magnetic field. This shows that FM interactions are still present in this systems, however, the overall behavior of the $M/H(T)$ is not that of a simple ferromagnet.

This scenario becomes more evident from the magnetization curves, $M(H)$, at 2\,K for both field directions, which are shown in Fig.~\ref{fig2} for the same three crystals as in Fig.~\ref{fig1}. For all three concentrations, $x=0.9$, $x=0.8$, and $y=0.9$, the magnetization is larger for $H\perp c$ (black curves) compared to $H\parallel c$ (red curves) with a tendency towards saturation at high fields. For the magnetically ordered systems, $x=0.8$ and $y=0.9$, no remanent magnetization at zero field is observed in the magnetically ordered phase, neither for $H\perp c$, nor for $H\parallel c$. This excludes a FM ground state in these materials. Comparing these data to the established ferromagnetic materials at $x=0$ (Fig.~4 in Ref.~\cite{Krellner:2008b}) and $y=0.3$ (Fig.~2b in Ref.~\cite{Jesche:2012}), one would have expected a finite remanent magnetization for $H\parallel c$. Instead, a linear $M(H)$ is observed for $0\leq \mu_0H\leq 0.6$\,T for $x=0.8$ and $0\leq \mu_0H\leq 0.3$\,T for $y=0.9$. In addition at higher magnetic fields a metamagnetic increase of the magnetization is found, which is enlarged in the insets of Fig.~\ref{fig2}b,c. Remarkably, these metamagnetic transition look very similar for the two different substitutions, with a small hysteresis loop, indicating a first-order type transition below 1\,T. In contrast to recent NMR studies \cite{Kitagawa:2011}, we do not find metamagnetic behavior up to $\mu_0H=7$\,T within the magnetically easy plane ($H\perp c$).   

\begin{figure}
\includegraphics[width=0.9\columnwidth]{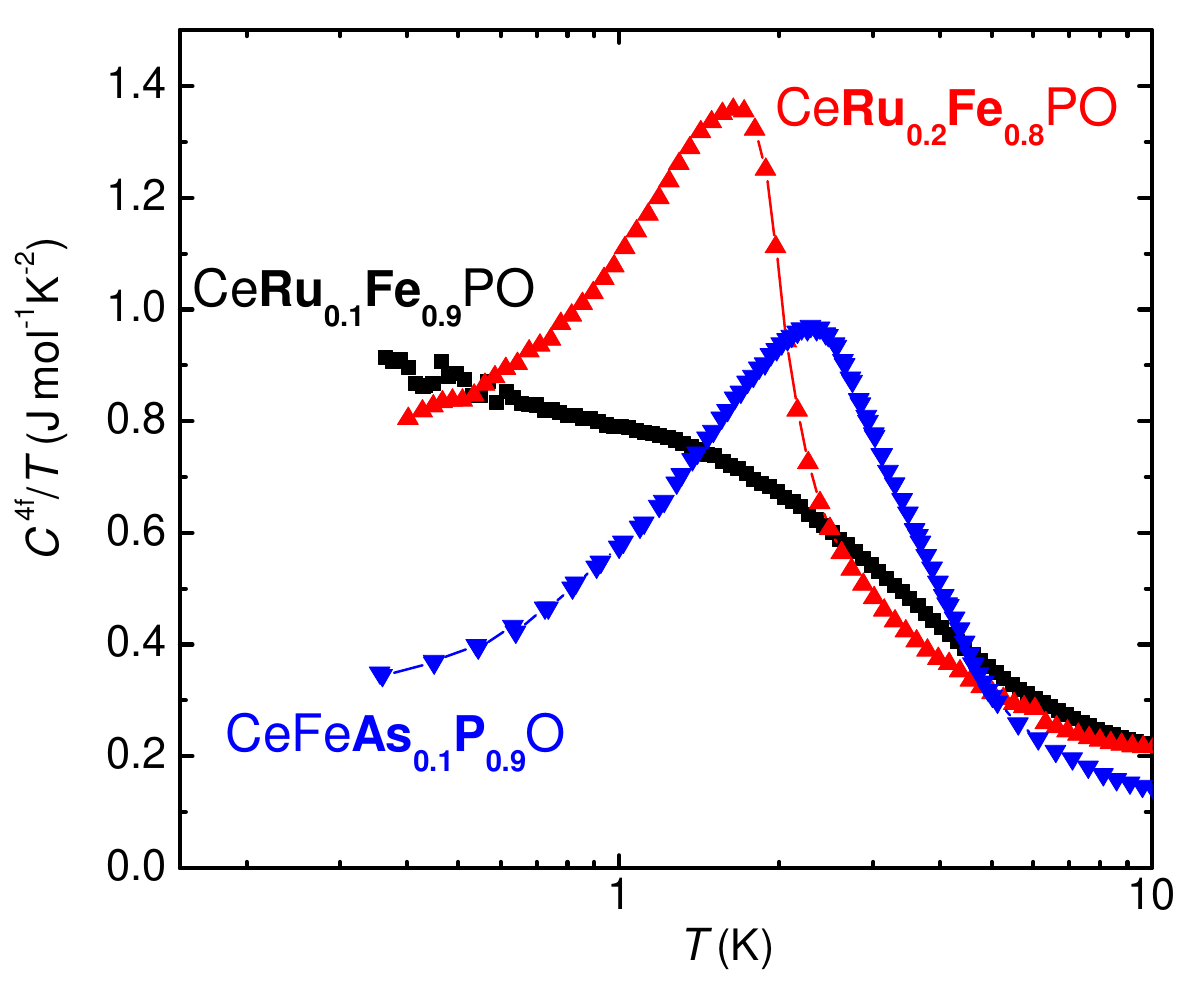}%
\caption{\label{fig3} (Color online) $4f$-contribution to the specific heat as function of temperature at zero magnetic field for samples with $x=0.9$ (squares), $x=0.8$ (up-pointing triangle), and $y=0.9$ (down-pointing triangle).}%
\end{figure}

The temperature dependence of the specific heat, $C^{4f}/T(T)$, for the three substitution levels is presented in Fig.~\ref{fig3}. For $x=0.9$, no phase transition is observed down to 0.4\,K, instead, $C^{4f}/T$ strongly increases with decreasing temperature, typical for a heavy-fermion system in the vicinity of a quantum-critical point. However, there is not a unique temperature dependence over more than a decade in temperature and it seems that the divergence gets weaker below 1\,K. Presently, we cannot exclude that the weakening of the divergence is due to short-range magnetic order at lower temperatures, which in case of single-crystalline CeFePO has lead to a broad hump, centered around $T_g\sim0.5$\,K \cite{Lausberg:2012}. The application of a magnetic field leads to a saturation of $C^{4f}/T$ at low $T$. For $\mu_0H=5$\,T a constant Sommerfeld coefficient, $C^{4f}_{5T}/T=0.6$\,Jmol$^{-1}$K$^{-2}$, is observed below 1\,K, which decreases further with increasing field to $C^{4f}_{5T}/T=0.4$\,Jmol$^{-1}$K$^{-2}$ for $\mu_0H=9$\,T. 

For $x=0.8$ and $y=0.9$, the specific-heat data in zero magnetic field present a large anomaly at the magnetic phase transition. For $y=0.9$ the transition takes place at a slightly higher temperature  compared to the $x=0.8$ sample in agreement with the susceptibility data. The large anomaly proves that the magnetic phase transition is due to long-range magnetic order. This is further corroborated by an analysis of the magnetic entropy, obtained by integrating $C^{4f}/T$ over temperature. The increase in entropy from $T=0.4$\,K to $5$\,K amounts to $0.56R\ln2$ for $x=0.8$, and $0.52R\ln 2$ for $y=0.9$, respectively. These values are well above the observed entropy gain for the short-range ordered state in CeFePO, which was only $0.01R\ln2$, but still below the expected $R\ln2$ due to a pronounced Kondo screening.  

\begin{figure}
\includegraphics[width=\columnwidth]{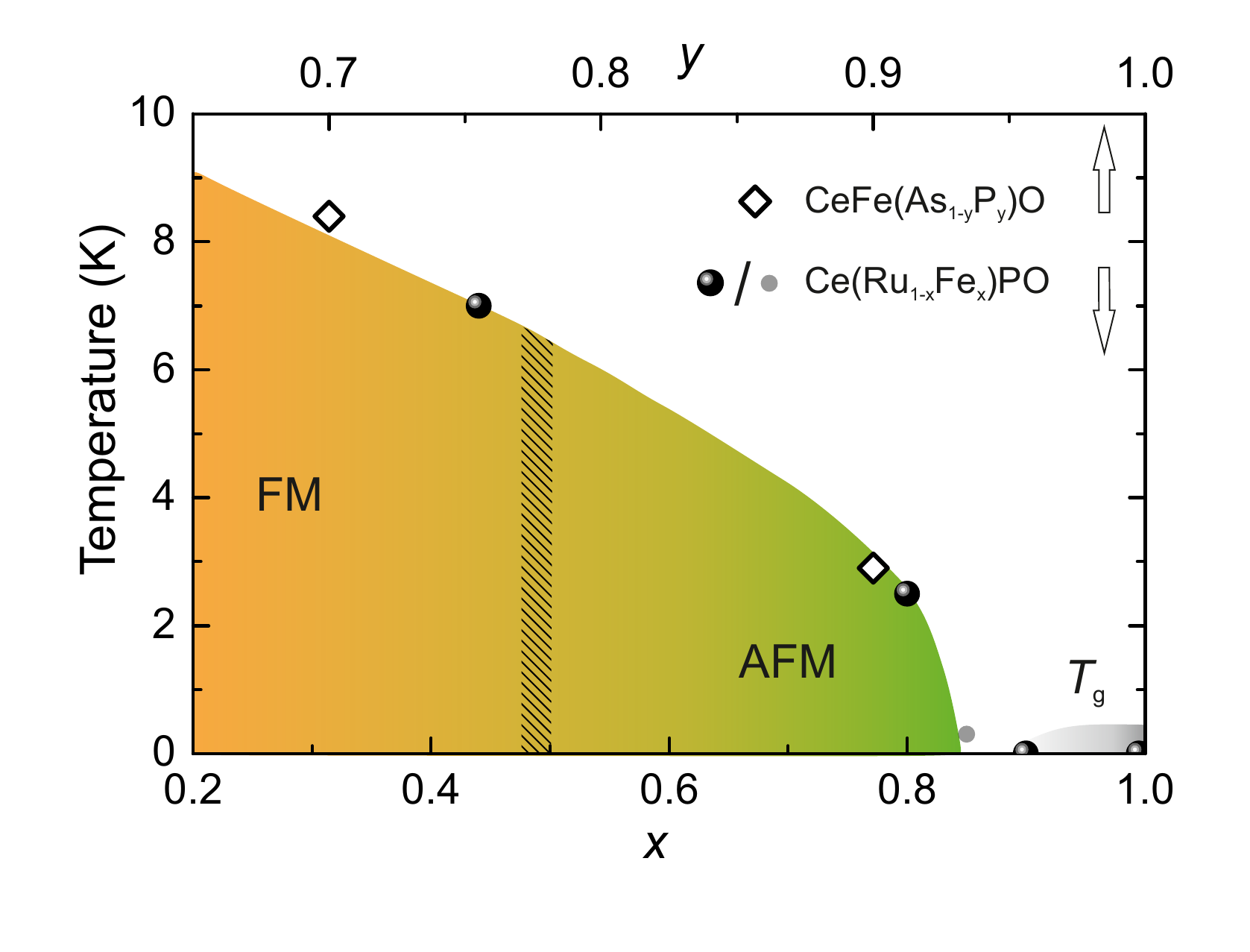}%
\caption{\label{fig4} (Color online) Temperature-substitution phase diagram of the series \Ru (circles, lower abscissa) and \As (diamonds, upper abscissa) obtained from magnetic measurements. The small gray point at $x=0.85$ was taken from Ref.~\cite{Kitagawa:2012}. A transition from ferromagnetism to antiferromagnetism indicated by the shaded bar occurs before reaching the quantum critical point at $x=0.85$.}
\end{figure}

The obtained results are summarized in a temperature-concentration phase diagram presented in Fig.~\ref{fig4}, which is markedly different to the reported phase diagram in Ref.~\cite{Kitagawa:2012} because we have found clear evidence that the ferromagnetic transition changes into an antiferromagnetic one, when approaching the quantum critical point. Presently, we cannot put an exact boundary of the FM to AFM transition, therefore a shaded color code with an indicative line was used to show, that at some concentration this transition takes place. A similar change from ferromagnetism to antiferromagnetism was also observed in single crystalline CeRuPO samples under pressure. There, the transition occurs at $p^*\sim 0.87$\,GPa \cite{Lengyel:2015}. This pressure corresponds to a relative volume change of 0.83\% using the experimental bulk modulus of $K_0=105(1)$\,GPa, determined by X-ray diffraction under pressure \cite{Hirai:2013}. Assuming Vegard's law for the lattice volume of the series \Ru, such a volume change would be achieved already for $x=0.2$, but experimentally we still observe the ferromagnetic ground state for $x=0.44$. However, the lattice parameters $a$ and $c$ in the series \Ru do not evolve equally, $a$ decreases with increasing $x$, whereas $c$ increases with $x$ \cite{Kitagawa:2012}. Therefore, chemically induced pressure is expected to be different compared to hydrostatic pressure, as the $c/a$ ratio develops differently. This might be also the reason why in the pressurized CeRuPO no QCP could be reached, which is clearly the case for \Ru.

\section{Conclusion}
We have succeeded in growing single crystals of the two substitution series \Ru 
\linebreak and \As, which are large enough to determine the magnetic ground state by means of magnetic-susceptibility and specific-heat measurements. We clearly demonstrate that the ferromagnetic ground state at $x=0.44$ and $y=0.7$ changes into an antiferromagnetic one, when approaching the quantum critical point. This is manifested by the absence of a remanent magnetization at zero field in the magnetically ordered state. However, ferromagnetic interactions are still present in these materials, reflected by pronounced metamagnetic transitions below 1\,T for $H\parallel c$, which is the magnetically hard direction. Our results verify that a ferromagnetic quantum critical point is avoided in substituted CeFePO, similar to what was observed in other two-dimensional heavy-fermion ferromagnets, when $T_C$ was tuned towards zero \cite{Brando:2016}. These results further corroborate that the key towards ferromagnetic quantum criticality, which was observed in YbNi$_4$P$_2$, might be the quasi-one-dimensional crystal and electronic structure present in that system. 

\begin{acknowledgement}
The authors thank U. Burkhardt and P. Scheppan for energy dispersive X-ray analysis of the samples as well as R. Weise and K.-D. Luther for technical assistance. 
This work was supported by the DFG priority program SPP 1458. 
\end{acknowledgement}


\bibliography{JescheARXIV}

\end{document}